\documentstyle[11pt]{article}
\def\li#1#2{#1{\sc\lowercase{#2}}} 
\def\OII{[\li{O}{II}]}
\def\gs{\mathrel{\raise0.3ex\hbox{$\scriptstyle >$}\kern-0.70em %
\lower0.71ex\hbox{{$\scriptstyle \sim$}}}}
\def\ls{\mathrel{\raise0.3ex\hbox{$\scriptstyle <$}\kern-0.70em 
\lower0.71ex\hbox{{$\scriptstyle \sim$}}}}
\def\et{\hbox{\it et al.}$\,$}
\def\go{
\mathrel{\raise.3ex\hbox{$>$}\mkern-14mu\lower0.6ex\hbox{$\sim$}}
}
\def\lo{
\mathrel{\raise.3ex\hbox{$<$}\mkern-14mu\lower0.6ex\hbox{$\sim$}}
}
\def\w{\omega(\theta)}
\def\A{A_\omega}
\def\li#1#2{#1{\sc\lowercase{#2}}}
\def\OII{[\li{O}{II}]}
\def\epsiras{\epsilon_{\rm IRAS}}

\begin{document}

\centerline{\LARGE {\bf Evolution in the Clustering of Galaxies to $r$ = 26}}
\vskip 0.8cm
\centerline{\large Tereasa G. Brainerd$^1$, Ian Smail$^2$, and Jeremy
Mould$^3$}
\vskip 0.3cm
\begin{centering}
\normalsize
$^1$Theoretical Astrophysics, Caltech 130-33, Pasadena, CA 91125, USA \\
$^2$Palomar Observatory, Caltech 105-24, Pasadena CA 91125, USA \\
$^3$Mount Stromlo and Siding Springs Observatories,
Private Bag, Weston P.O., ACT 2611, Australia \\
\end{centering}

\vskip 1.0cm
\centerline{ABSTRACT}
\vskip 0.3cm
{\normalsize
We present results for the  two-point angular correlation function of
galaxies, $\w$, to a limiting magnitude of $r = 26$.  Our catalogue is
constructed from deep imaging using the COSMIC imaging spectrograph on
the Hale 5-m.  The final sample is 97\% complete to $r=26.0$ yielding
$\sim$5700 galaxies over a 90.1 sq.\ arcmin field.  Our analysis shows
$\w$ for faint galaxies can be parameterised by a power law of the form
$\A \theta^{-0.8}$, in agreement with
the angular clustering statistics of shallower catalogues.  The derived
amplitude, $\A$, for our catalogue is small, but non-zero.  We
combine this measurement with the latest statistical constraints on
faint galaxy redshifts from gravitational lensing studies, which imply
that the bulk of the $r\ls26$ field galaxies should be at redshifts
$z\sim1$.  We show that our derived $\A$ is significantly lower than
that predicted from the local bright, optically-selected galaxy
correlation function using the lensing-determined galaxy redshift distribution
and modest growth of clustering. However, this simplistic model
does not include the variation in observed morphological mix as a
function of redshift and apparent magnitude in our sample.  At our
faintest limits we reach sufficiently high redshifts that differential
$K$-corrections will result in the observed galaxy mix  being dominated
by star bursting dwarf and low surface brightness irregulars, rather
than the early-type systems used to define the local bright galaxy
correlation function.   Adopting the correlation function measured
locally for these low surface brightness galaxies and assuming modest
clustering evolution, we obtain reasonable
agreement between our model and observations.
This model, therefore, supports the scenario in which the high
number density of faint galaxies is produced by normally clustered
star forming dwarf galaxies at modest redshifts.}


\vskip 1.0cm
\noindent{\bf 1 INTRODUCTION}
\vskip 0.25cm

The observed surface density of faint galaxies far exceeds the number
predicted from a simple extrapolation of the local universe (the no
evolution model; NE), irrespective of the cosmological geometry.  However, to
the faintest limits measurable the distribution of galaxies in redshift is
a close match to that predicted by the NE model.  Since at fainter limits
galaxies become progressively bluer, the `excess'
population is primarily caused by an increase in the space density
of blue low-luminosity galaxies at modest redshifts.  The exact nature
of this blue excess population is currently one of the major questions
in observational cosmology and has significant impact on our
understanding of galaxy formation and evolution.

Observationally, one of the  cheapest statistics to obtain for samples
of faint galaxies is the strength of their clustering as seen in
projection on the plane of the sky.  This is commonly measured using
the two-point angular correlation function, $\w$, defined by
\begin{equation}
dP = n^2 [1 + \w] d\Omega_1 d\Omega_2,
\end{equation}
where $n$ is the mean number density of galaxies and $dP$ is the
probability in excess of Poisson of observing a pair of galaxies in
solid angle elements $d\Omega_1$ and $d\Omega_2$ that are separated by
an angle $\theta$ on the sky. To a given depth, $\w$ is dependent upon
the redshift distribution of the sources, $N(z)$, and their
3-dimensional real space clustering strength, $\xi(r,z)$.  Thus $\w$ is
a sensitive test of both the luminosity and clustering evolution of
faint galaxies (Koo \& Szalay 1984). This fact, combined with the
advent of efficient large format CCD detectors, has led to a recent
flurry of interest in using the correlation strength of faint galaxies
to study their nature (Neuschaefer, Windhorst \& Dressler 1991;
Efstathiou \et 1991; Couch, Jurcevic \& Boyle 1993; Roche \et 1993).

At the bright magnitudes accessible with plate material from Schmidt
telescopes ($B\ls20$), $\w$ has been shown to follow a power law form:
$\omega(\theta) = A_\omega \theta^{-\delta}$ with $\delta \approx 0.8$
(Groth \& Peebles 1977; Shanks \et 1980; Maddox \et 1990).  The value
of $A_\omega$ depends upon the depth of the survey, but the derived
values are in reasonable agreement when the results from different
catalogues are rescaled to similar depths.  Using plates from 4-m class
telescopes it is possible to go somewhat fainter ($B\ls24$), while
still covering a large area.  The studies at this depth are in less
agreement, but do seem to support a continued $\delta \approx 0.8$
exponent for $\w$ to at least $B\sim23$ (Koo \& Szalay 1984; Stevenson
\et 1985 (SSFM)).

To reach still fainter limits CCDs have to be employed, with a
resulting large reduction in sky coverage.  As mentioned, several
groups have attempted to measure $\w$ using CCD imaging and we briefly
summarise these studies.  Neuschaefer, Windhorst \& Dressler (NWD,
1991) used $g$ imaging over a 720 sq.\ arcmin area complete to $g \sim
24.5$ (roughly $R \sim 23.5$) and claim little change in the index,
$\delta$, of $\w$ from the value at brighter magnitudes.  However, they
do see a strong decline in the amplitude, $A_\omega$, to $g\sim24.5$.
Efstathiou \et (EBKTG, 1991) used multi-colour data on a number of
small fields (each about 10 sq.\ arcmin) to measure $\w$ to $B < 26$ or
$R<25$.  Their sample shows very weak clustering at their faintest
limits and from this they conclude that a large fraction of the faint
galaxies in their sample belong to a weakly clustered population which
is not seen today.  By far the largest study is that of Couch, Jurcevic
\& Boyle (CSB, 1993) who analysed a 4 sq.\ degree area to $R\ls23$ and,
as NWD, they found no change in $\delta$ with depth, while $A_\omega$
declined strongly. They conclude that their observations are
inconsistent with linear growth in a standard CDM cosmogony.  Finally,
Roche \et (RSMF, 1993) analysed a total of roughly 300 sq.\ arcmin in
$B$ and $R$, with the $R$ sample  limited at $R\leq23.5$.  Adopting a
$\theta^{-0.8}$ power law for $\w$, they studied the variation of
$A_\omega$ with depth and their analysis shows the, by now familiar,
decline in $A_\omega$ at fainter magnitudes.  They interpret their
result as support for a model in which the faint galaxies have a
clustering amplitude similar to local bright galaxies,
but are distributed across a wide redshift range:
$1\lo z\lo 3$.

The recent study of Bernstein \et (1994) used both plate material and
statistical information on the redshift distribution of galaxies to
$B\sim22$ to estimate $\xi(r)$ on small scales at two different epochs:
$z\sim0.18$ and $z\sim0.27$.  Their results are significantly lower
than predictions of models extrapolated from local bright galaxy clustering.
However, their observations are consistent with the assumptions of modest
clustering growth and that
the bluest 60-70\% of their sample have
present-day clustering similar to IRAS-selected samples, which are
intrinsically more weakly clustered than  bright optically-selected samples
(Saunders, Rowan-Robinson \& Lawrence 1992, SRRL; Fisher \et 1994).

A more direct approach to study  the evolution of the two-point spatial
correlation function, $\xi(r)$, uses the extensive field redshift samples
which are gradually becoming available. From the large Broadhurst,
Ellis \& Colless AAT redshift survey, Cole \et (1994) analysed both the
evolution in $\xi(r)$ and possible differences between the clustering
of strongly and weakly star-forming galaxies (as indicated by their
\OII\ emission).  They detected no evolution in $\xi(r)$ out to
$z\sim0.6$ and no significant difference between the clustering
strengths of the \OII-weak and \OII-strong populations.  However, at the
limit of this study ($B\sim22$) neither the apparent surface density of
galaxies nor their clustering amplitude depart significantly from the
NE prediction and, thus, the measurements are not particularly
sensitive to any excess population.

The CCD studies of the angular correlation function of faint galaxies, while in
broad agreement about the strong decline of the clustering strength of
galaxies at fainter apparent magnitudes, differ in the interpretation
of this result. Their inconclusive nature has partly arisen from our
lack of information about the redshift distribution of very faint field
galaxies.  Most of the quantitative analyses undertaken adopt
pure-luminosity evolution models to describe $N(z)$ for the faint
galaxy samples.  These models, even in their milder forms, are in
conflict with the observed redshift distribution in faint field surveys
(e.g.\ Glazebrook \et 1994) and their applicability at even fainter magnitudes
must, therefore, be in doubt.  At the depths probed by the current samples
use to estimate $\w$,
it is likely that even 10-m class telescopes will be unable to
provide complete spectroscopic samples.  Nevertheless, some distance
information is available for  these very faint sources.

Firstly, spectroscopic redshifts have been measured for a small, but growing,
sample of giant gravitationally lensed arcs.  These  arcs are highly
distorted images of serendipitously placed galaxies seen through the
cores of rich, moderate redshift clusters.  They should, therefore,
represent an unbiased sample of high-redshift faint field galaxies.
This has been confirmed by Smail \et (1993), who determined optical and
optical-infrared colours of arcs and found that the colours were
representative of the bulk of the faint field population.  The current sample
of
giant arcs has a modest median redshift of $\left< z \right> \sim 1$
with intrinsic source magnitudes of $B\sim25$--26.

Statistical information is available on the distances of samples much larger
than that of the giant arcs from analyses of weak lensing by rich clusters
of galaxies (Smail \et 1994; Kneib \et 1994).  Both of these studies
indicate that the bulk of the $B\sim26$--27 population lies at
redshifts $z\sim1$.  In particular, the analysis of an $I$ selected
sample by Smail \et (1994) gives a preferred redshift distribution
peaked around $\left< z \right> \sim 0.8$, similar to the distribution
of the giant arc sample.  The general shape of the redshift
distributions indicated by the various lensing analyses are close to
their respective NE model predictions.  While the NE model is
physically implausible at the distances and, hence, look-back times
probed at these faint limits, the form of the predicted distribution is
similar to that expected from  the Burst model of Broadhurst,
Ellis \& Shanks (1988) (see also Babul \& Rees 1992).  This model
effectively steepens the faint end slope at moderate redshifts by
invoking short periods of intense star formation in intrinsically faint
galaxies\footnote{In the discussion that follows we will use the term
`dwarf' for these intrinsically low-luminosity and low surface
brightness (LSB) galaxies; this is used with no implication for the
morphological nature of the systems.} to boost the faint number counts,
while retaining the general form of the NE redshift distribution (see
also McGaugh 1994). The faint counts are thus dominated by a population
of blue, star-forming dwarf galaxies at modest redshifts.   We
therefore adopt the form of the NE redshift distribution as the $N(z)$
for our standard model in the analysis below.

Another currently fashionable galaxy evolution model, which predicts an
$N(z)$ for the faint field population close to that observed, is the
Merger model of Broadhurst, Ellis \& Glazebrook (1992).  Unfortunately,
the clustering evolution of the merging galaxy population in this model
is likely to be complex and little theoretical work has been done on
predicting this evolution.  We will therefore not compare our
observations with this model in the discussion below.

RSMF claim that at their very faintest limits ($B \sim 25$)
$\w$ reaches a lower limit and then
flattens out.  Such a flattening, if it is confirmed, might indicate
the presence of a high-redshift cut-off in the distributions of faint
galaxies, either due to galaxy formation or the Lyman limit entering
the bandpass for high-redshift sources.  The existence of such a large
population of  high-redshift galaxies would not be consistent with the
lensing-derived $N(z)$.  The simplest method to distinguish between
these possibilities is to repeat this measurement in a redder
passband.  Therefore, we have undertaken a study of the evolution of
the clustering strength of galaxies in a sample of faint galaxies
selected in a red passband (Gunn $r$) to an effective magnitude limit
deeper than all previous studies ($r\sim26$).  By combining these
observations with the latest constraints on the distance to the faint
galaxy population from lensing studies, we hope to gain new insight
into the clustering evolution of distant galaxies.  Our dataset also
differs from those used previously in that it has both good seeing and
high spatial sampling, resulting in negligible confusion at our
faintest limits.

\vskip 0.5cm
\noindent{\bf 2 OBSERVATIONS}
\vskip 0.25cm

The observations analysed here were originally taken to select targets
for a deep spectroscopic survey with the 10-m Keck-I telescope.  The
data have also been used to study the coherent distortion of faint
galaxy images resulting from weak gravitational lensing by large-scale
structure (Mould \et 1994).  The dataset and its reduction to a
catalogue of detected objects is detailed in Mould \et (1994) and here
we present only a brief outline.

The final dataset consists of a total of 24.0 ksec integration in Gunn
$r$ and 6.0 ksec in Gunn $g$ on a single blank field. This was all
taken under good conditions (seeing 0.7--0.9 arcsec) with the COSMIC
imaging spectrograph  on the 5-m Hale telescope. In direct imaging
mode, COSMIC has a 9.5$\times$9.5 arcmin field with 0.28 arcsec/pixel
sampling.  The final stacked $r$ image has a 1$\sigma$ surface
brightness limit of $\mu_r = 28.8$ mag arcsec$^{-2}$, seeing of 0.87
arcsec FWHM, and a total area of 90.1 sq.\ arcmin.   The object
catalogue created from this frame using the FOCAS image analysis
package (Valdes 1982) contains $\sim$6600 objects  brighter than the
80\% completeness limit of $r=26.2$.  Adopting an extremely
conservative magnitude limit of $r\leq26.0$, where the detections are
roughly 97\% complete, we obtain a cumulative surface density of 71.8
galaxies per sq.\ arcmin (5--7 times the NE prediction, depending upon
$\Omega_0$).  To calculate the equivalent $R$ limit we use the typical
galaxy colour at this depth and the photometric conversion of Kent
(1985) giving $R \sim r - 0.55$ and a limit of $R\ls25.5$.  The
$g$-band data are deep enough to provide $g\! - \! r$ colours accurate
to $\Delta(g\! - \! r) \sim 0.2$ at this limit.  Our magnitude
definition follows that of Lilly \et (1991), adopting isophotal
magnitudes until an object's isophotal diameter shrinks below 3 arcsec,
at which point photometry within a fixed 3 arcsec aperture is used.
The catalogue of galaxies with $r\leq26.0$ is used as the basic data
for the analysis detailed below.

A mask frame defining areas around bright objects and along the frame
boundaries, where the  efficiency of galaxy detection is lower than
average, was also constructed.  Approximately 5\% of the total area of
the frame is masked out and these regions are excluded in the analysis
below.

\vskip 0.5cm
\noindent{\bf 3 ANALYSIS}
\vskip 0.25cm
\noindent{\bf 3.1 Estimation of $\w$}
\vskip 0.25cm

We estimate $\w$ for the galaxies using the direct pair-counting method
proposed by Landy \& Szalay (1993),
\begin{equation}
\hat{\omega}(\theta) = \frac{DD - 2DR + RR}{RR},
\end{equation}
where $DD$, $DR$, and $RR$ are the number of distinct data-data,
data-random, and random-random pairs (appropriately scaled by the
number of data and random points) in a given angular separation bin.
To estimate $\w$ we use 50,000 random points over the geometric area
covered by the data catalogue, taking into account the regions which
are masked out due to the presence of bright objects or the frame
edge.  The mean $\hat{\omega}(\theta)$ obtained from 50 independent
sets of random points is computed for angular scales, $17'' \le \theta
\le 427''$, valid for our dataset (0.14 to 3.5 Mpc at the median
redshift of our adopted $N(z)$).  Errors for $\hat{\omega}(\theta)$ are
estimated using 50 bootstrap resamplings of the data (e.g. Barrow,
Bhavsar, \& Sonoda 1984).

As a consistency check of $\hat{\omega}(\theta)$ obtained using the
direct estimator above, we have applied a counts-in-cells estimator
\begin{equation} \hat{\omega}(\theta)= \frac{\left<N_i
N_j\right>}{\left<N_i\right> \left<N_j\right>}-1, \end{equation} where
$N_i$ and $N_j$ are the number of galaxies found in cells $i$ and $j$,
and the brackets denote an average over pairs of cells separated by an
angle $\theta \pm \delta \theta$.  The galaxies were binned in $5''
\times 5''$ cells and $\hat{\omega}(\theta)$ computed using
equation~(3) above.  Excellent agreement was found between the results
of the direct and counts-in-cells estimators.

\vskip 0.25cm
\noindent{\bf 3.2 $\A$ and the Integral Constraint}
\vskip 0.25cm

In the calculation of $\hat{\omega}(\theta)$ the observed number
density of galaxies on the CCD frame (to a given magnitude limit) is
used to estimate the true mean density of galaxies at that magnitude
limit.  The small area of the frame results in a bias in the estimate
of $\w$ due to the well-known ``integral constraint'' which has the
effect of reducing $\hat{\omega}(\theta)$ by an amount
\begin{equation}
C = {1 \over \Omega^2} \int \int \w d\Omega_1 d\Omega_2,
\end{equation}
where the integrals are performed over the total solid angle, $\Omega$,
of the regions of the field not excluded by the detection
mask.

Assuming a power law form for the two-point correlation function, $\w =
\A \theta^{-\delta}$ with $\delta = 0.8$, as suggested by previous
brighter surveys, we find $C = 0.0146 \A$ for our field geometry and
galaxy detection mask.

In order to compare estimates of $\w$ from catalogues with differing
depths, it is common to quote the value of $\hat{\omega}(\theta =
1^\circ)$ after adopting a power law form for $\hat{\omega}(\theta)$
with $\delta = 0.8$ and correcting $\A$ for both the integral
constraint and the dilution effect of faint stars which have not been
removed from the object catalogues.  The corrected amplitude is then
\begin{equation}
\A^{\rm corr} = \left( {N_{\rm obj} \over N_{\rm obj} - N_{\rm s}}
                \right)^2 \A^{\rm IC},
\end{equation}
where $N_{\rm obj}$ is the total number of objects used in the
determination of $\hat{\omega}(\theta)$, $N_{\rm s}$ is the number of
contaminating stars (estimated from the model of Bahcall \& Soneira
1980), and $\A^{\rm IC}$ is the amplitude of the best-fit power law
with $\delta = 0.8$, corrected for the
integral constraint.  Table 1 summarizes our results for these
quantities, where the value of $\A^{\rm corr}$ is appropriate for
$\theta$ in units of arcseconds.
The error on $\A^{\rm corr}$ has been estimated from Monte
Carlo simulations of fields populated with galaxies distributed with a
known $\omega(\theta)$ of the form $\A\theta^{-0.8}$.  These simulated
fields were constructed using the iterative tree method of Soneira \&
Peebles (1978).  The resulting fractional error in $\A$ rises from 10\%
for our brightest bin to 20\% for the faintest.  The stellar contamination
correction from the Bahcall \& Soneira (1980) model is highly uncertain;
however, a factor of 2 increase the number of stars in our faintest
bins amounts to only a 1 $\sigma$ increase in $\A$.

\newpage
\noindent{\bf 3.3 Expected Correlations}
\vskip 0.25cm

{}From local bright, optically-selected galaxy surveys, the two-point
spatial correlation function, $\xi(r)$, is well-approximated by a power
law of the form $\xi(r) = (r/r_0)^{-\gamma}$, $\gamma \approx 1.8$ and
$r_0 \approx 5.5 h^{-1} {\rm Mpc}$, for $10 h^{-1} {\rm kpc} \lo r \lo
10 h^{-1} {\rm Mpc}$ ($h$ is the present value of the Hubble parameter
in units of 100 kms sec$^{-1}$ Mpc$^{-1}$).  The angular correlation
function is related to the spatial correlation function by an integral
equation (see, for example, Peebles 1980).  Parameterizing the
evolution of $\xi(r)$ by
\begin{equation}
\xi(r,z) = \left( {r \over r_0} \right)^{-\gamma}
           \left(1+z \right)^{- \left( 3+\epsilon \right)},
\end{equation}
the relation between $\w$ and $\xi(r,z)$ for small angles is
\begin{equation}
\w = \sqrt{\pi} {\Gamma[(\gamma-1)/2] \over \Gamma(\gamma/2)}
     {A \over \theta^{\gamma-1}} r_0^\gamma
\end{equation}
where
$$ A = \int_0^\infty g(z) \left( {dN \over dz} \right)^2 dz
      \left[ \int_0^\infty \left( {dN \over dz} \right) dz \right]^{-2}$$
and
$$ g(z) = \left( {dz \over dx} \right) x^{1-\gamma} F(x)
          \left( 1+z \right)^{-(3 + \epsilon - \gamma)}.$$
Here $x$ is the coordinate distance at redshift $z$, $r$ is the proper
coordinate, $dN/dz$ is the number of galaxies per unit redshift, and
the metric is
$$ ds^2 = c^2 dt^2 - a^2[dx^2/F(x)^2 + x^2 d\theta^2 + x^2 {\rm sin}^2 \theta
          d\phi^2].$$
For a power law parameterisation with constant $\gamma$,
the evolution of $\xi(r)$ with redshift is given by $\epsilon$.  Assuming
$\gamma = 1.8$, linear
theory predicts $\epsilon = 0.8$, while clustering
fixed in proper coordinates yields $\epsilon = 0.0$, and clustering
fixed in comoving coordinates yields $\epsilon = -1.2$.

The small-scale clustering of galaxies is a highly non-linear process,
and it is not at all obvious what value of $\epsilon$ to choose in
order to compare model predictions of $\w$ to observations. Neither is
it apparent that the assumption of a unique power law form for
$\xi(r,z)$ is appropriate.  Relatively few numerical studies of the
evolution of $\xi(r)$ from redshifts of order $z\sim3$--4 to the
present have been performed.  Yoshii, Peterson, \& Takahara (1993)
investigated the evolution of $\xi(r)$ of ``density peak tracers'' in
N-body simulations of cold dark matter (CDM) universes from $z=2$ to
$z=0$ and concluded that a flat CDM model is in agreement with the low
clustering amplitude of faint galaxies observed by EBKTG.

Brainerd \& Villumsen (1994) used a large N-body simulation to
investigate the evolution of $\xi(r)$ of individual ``dark matter
halos'' in a flat CDM universe from $z=5$ to $z=0$ and found that the
evolution of $\xi(r)$ for these objects depended strongly on the
parameters used to classify groups of particles as ``halos'' (eg. mass
and overdensity).  The correlation functions of moderate overdensity
halos ($\delta\rho/\rho \sim 250$) and high-mass, high-overdensity
halos ($\delta\rho/\rho \sim 2000$) were well-described by a power law
whose index, $\gamma$, remained constant over the course of the
simulation. However, $\gamma$ for low-mass, high-overdensity halos
increased continuously.  In all cases, the ``first generation'' of
halos was highly clustered, in agreement with the predictions of
biased galaxy formation (eg. Bardeen \et 1986).
Over the epoch of halo formation, however,
the distribution of moderate overdensity halos became {\it less}
clustered than that of the first generation, while the very overdense halos
remained at the same clustering strength.
For those halos whose $\xi(r)$ was well described by a unique $\gamma$,
two general patterns in the evolution of $\xi(r)$ were observed after
all halos had formed:  $\epsilon = 0.0 \pm 0.2$ (low-mass,
moderate-overdensity halos) and $\epsilon = -1.2$ (high-mass, moderate
overdensity halos and high-mass, high-overdensity halos).  In the following
analysis we adopt these values of $\epsilon$,
along with the linear theory prediction $\epsilon = 0.8$,
as representative of likely evolution in $\xi(r)$.  We do note, however,
that more rapid rates of clustering evolution have been observed in
N-body simulations (eg.\ Melott 1992, Davis \et 1985).

\vskip 0.5cm
\noindent{\bf 4 RESULTS AND DISCUSSION}
\vskip 0.25cm

Figure~1 shows $\hat{\omega}(\theta)$
for our three faintest samples together with the best-fit power laws,
$\A \theta^{-\delta}$, all of which have indices of $\delta \sim 0.8$.  This
indicates
that the $\theta^{-0.8}$ power law form for $\hat{\omega}(\theta)$ measured at
bright magnitudes continues all the way to our faintest sample.  As observed
by previous investigators, the amplitude of $\hat{\omega}(\theta)$ decreases
with the effective depth of the catalogue.  We illustrate this with Figure~2,
in which we have plotted $\hat{\omega}(\theta)$ extrapolated to $\theta =
1^\circ$ (using $\hat{\omega}(\theta) \propto \theta^{-0.8}$) as a
function of limiting $R$ magnitude for a number of red-selected
samples.  It is readily apparent that our results extend the trend of
weaker correlation at fainter depths, with no indication of any
flattening of this decline to our faintest limits.  A weighted linear
least squares fit to the data in Figure 2 yields $\hat{\omega}(\theta =
1^\circ) \propto R^{-0.27 \pm 0.01}$.  While these results are
from only a single field, they are in qualitative agreement with
preliminary results from a larger
study by Bernstein \& collaborators (Bernstein, priv.\ comm.).

The distribution of object colours on our frame radically
changes at $r\sim23$, when the median colour shifts from $g\! - \! r \sim 1$
to $g\! - \! r \sim 0.1$--0.2.
Using these $g\! - \! r$ colours we, therefore, split the sample by colour
at $g\! - \! r = 0.3$ and determine the relative clustering strengths of
the ``red'' and ``blue'' galaxies.  To
within the measurement errors, there  is no apparent
difference in $\hat{\omega}(\theta)$ between the two subsamples.

We compare our observed amplitudes of $\hat{\omega}(\theta)$ with that
predicted by our standard model, a single population of galaxies which
have a present-day correlation length typical of local bright galaxy
samples.  For simplicity this population is assumed to have the same mix of
galaxy types visible at each epoch.
To make this comparison we calculate the appropriate NE
$N(z)$ for a surface brightness selected sample of
galaxies in each of our magnitude intervals (King \& Ellis 1985) and, for
convenience, we parameterise the shape of these predicted redshift
distributions by (EBKTG)
\begin{equation}
{dN \over dz} \propto
z^2 \exp{\left[-\left({z\over z_0} \right)^\beta \right]}.
\end{equation}
(It should be noted that the predicted correlation amplitude is dependent
upon only the shape of $N(z)$ and not its normalisation.)
Adopting $\gamma = 1.8$, $r_0 = 5.5h^{-1}$ Mpc, and choosing a value of
$\epsilon$, we calculate $\A$ for different magnitude ranges using
equation~(7) above.  The model parameters $z_0$ and $\beta$ are
summarized in Table~2 for the cases of (1) a flat universe with
$\Omega_0 = 1$ and (2) an open universe with $\Omega_0 = 0.2$.

Figure 3 shows our observed variation in $A_\omega^{\rm corr}$ with
limiting magnitude and model predictions using $N(z)$ from Table~2 with
the choices of $\epsilon=-1.2$, 0.0, 0.8 and $\Omega_0 = 1.0$, 0.2.
These models predict an amplitude that is about an order of
magnitude greater than that observed.  Clearly at least one of the
assumptions in the models is incorrect.  We first investigate
the individual variation in each of the main model parameters ($\epsilon$,
$N(z)$, and $r_0$) necessary to match our observations.

If we allow only $\epsilon$ to
vary in the models, then we find it must be of order 6 or 7 for the
predicted amplitude to match that observed ($\epsilon_{\rm stan}$ in
Table~2), implying a
very rapid evolution in the correlation function: $\xi(z) \sim
(1+z)^{-7}$.  Such extreme evolution in the clustering of the
{\it entire} population
would have been easily detected  by Cole \et (1994) and, so, we prefer
to search for another explanation.

One possibility is to relax our limits on the redshift
distributions.  By distributing the galaxies over both a wider range of
redshifts and out to higher redshifts, it is possible to weaken the
predicted clustering amplitude substantially.  Using the measured
correlation amplitude and assuming $\epsilon = 0.8$, we determine the
minimum values of $z_0$ and $\beta$ that are consistent with the 95\%
confidence level upper limit on the measured amplitude.  This choice of
$\epsilon$ allows for clustering that evolves at the rate predicted by
linear theory and, amongst our 3 standard choices for $\epsilon$,
yields the lowest amplitude.  To further constrain the possible
distributions, in particular the value of $\beta$, we require that
$N(z)$ comply with the limits on the fraction of galaxies with $z>3$
from the study of Guhathakurta, Tyson \& Majewski (1990).  The minimum
allowable parameters of the model $N(z)$ are given in Table~3 and indicate
that by putting the galaxies at a higher median redshift, the growth of
$\xi(r)$ required to match our observed amplitude of
$\hat{\omega}(\theta)$ is much less extreme than that predicted using the
$N(z)$ suggested by the lensing studies.  However, to achieve the necessary
dilution of the clustering strength we require the median redshifts in
our sample to be roughly 3--4 times the values obtained from the lensing
analyses.  This seems unlikely, although such a distribution might be
compatible if the scale sizes of faint galaxies decrease very strongly
with redshift.

In our standard model we made the simplifying assumption that at each
epoch we would observe the same mix of morphological types.   This is
not valid in our passband at the depths probed, even for the NE model.
The 4000 \AA $\;$ break moves through the $r$-band at a redshift of $z
\sim 0.7$, and this is the median redshift proposed for our faintest
samples.  The presence of strong 4000 \AA $\;$ breaks in early-type
galaxies compared to late-types, in the absence of luminosity
evolution, causes our deepest samples to be dominated by late-type
spiral and irregular galaxies.  At $r\sim20$ the mix of types in the NE
model is (E/S0/Sab, Sbc/Scd/Sdm):  (98\%, 2\%).  However, by $r\sim26$
this has changed to (37\%, 63\%).  In the Burst model this shift is
even more extreme, as the `excess' population in the faint counts
consists exclusively of bursting gas-rich late-type spirals.  This
preeminence of moderate-redshift late-type systems in our catalogue is
supported by the observed colour distribution, which peaks at $g\! - \!
r \sim 0.1$--0.2, typical of late-type spirals and irregulars at
$z\sim1$.  If these late-type galaxies are significantly less clustered
than our adopted local sample, we would observe an amplitude for the
correlation function which gradually falls below the predictions of our
standard model at fainter magnitudes.

There exist local late-type galaxy populations
that exhibit weaker clustering than the bright optical galaxy
samples used in this comparison (e.g.\ dwarf galaxies, LSB galaxies,
IRAS galaxies).  These could serve as a
natural mechanism for diluting the clustering strength of the faint
galaxy samples through a progressive domination by weakly clustered,
late-type dwarf galaxies.  We can predict the
necessary correlation length of these galaxies must have to agree with our
observations.  Summarised in Table~4 are the correlation lengths,
$r_0$, required of this population in order to match the 95\%
confidence upper limits on our observed amplitudes.  The choice of an
open or closed universe makes little difference in the required
correlation length, and for all three values of $\epsilon$, $r_0$ is
less than half the present-day correlation length of bright,
optically-selected galaxies.

By a similar argument Bernstein \et (1994) were able to construct a
viable model for their observations by adopting the correlation function of
IRAS galaxies for the dominant blue population in their $B\sim22$ sample.
They then required only modest clustering
evolution ($\epsilon \sim 0$) for the bulk
of their $z\sim0.3$ population.  Shown in Table~2 is the clustering evolution
needed to reproduce our values of $\A$ using a population described by
the IRAS correlation function of SRRL ($\gamma \sim -1.6$ and $r_0 \sim
3.8 h^{-1}$ Mpc).  Under these assumptions our correlation function has to
evolve as $\epsiras \sim 3$, which is slower than our standard
model predicts, but considerably faster than the theoretical predictions.

We would claim that the IRAS correlation function is probably not the
correct choice to describe the faint galaxy population.  A more
representative local sample would be either dwarf or LSB galaxies (see
McGaugh 1994), especially within the framework of the Burst model
described above.  Two recent, large studies have measured correlation
functions for similar classes of galaxies.  Santiago \& da Costa (1990)
used volume-limited samples from the new Southern Sky Redshift Survey
(SSRS) to study the clustering of LSBs, while Thuan \et (1991) analysed
a mixed sample of dwarf and LSB galaxies from the UGC.  The more
interesting of these is the  SSRS study which finds a correlation
length for LSB galaxies of $r_0 \sim 2.3$--2.7$ h^{-1}$ Mpc, depending
upon the exact sample definition.  These local values are
somewhat uncertain due
to the small samples available, however, the range of values is in
reasonable agreement with those quoted in Table~4 for clustering growth
of $\epsilon \sim 0.8$.  It therefore appears that we can achieve a
sufficiently low amplitude for $\w$ at faint limits by postulating a
gradual transition to LSB, late-type dominated samples via differential
$K$-corrections, coupled with the intrinsically weaker clustering of
these galaxies.

\vskip 0.5cm
\noindent{\bf 5 CONCLUSIONS}
\vskip 0.25cm

We have measured the angular correlation function of galaxies to an apparent
magnitude limit of $r\leq26$.  The correlation amplitude, $\A$, for our
faintest sample is extremely small, but significantly
non-zero.  This detection is
approximately an order of magnitude weaker than the predictions of a
theoretically and observationally motivated standard model.  The
amplitude of the correlation function decreases continuously to our
faintest limits, showing no evidence for any flattening (c.f.\ RSMF).
We find no difference between the clustering strength of the blue and
red galaxies within our sample.

There are three main parameters in our standard model which can be
varied in order to obtain agreement with the observations: the
evolution in $\xi(r)$ (parameterised by $\epsilon$), the redshift
distribution of the sample, $N(z)$, and  the correlation length,
$r_0$.  Forcing the observed low correlation amplitude to result from a
variation of either $\epsilon$ or $N(z)$ results in two
unpalatable solutions: extremely rapid (and unexpected) evolution in
$\xi(r)$ or a redshift distribution in which the galaxies are distributed
at median redshifts 3--4 times that expected from gravitational lensing
studies of rich clusters.

We prefer a more natural explanation for
the low clustering strength observed, resulting from the tendency of
deep field samples to be dominated by moderate-redshift,
blue star-forming spiral and
irregular galaxies.  This arises from the different $K$-corrections
for early- and late-type systems.  However, this effect is magnified by
the overwhelming prevalence of blue late-type systems in faint field counts,
as illustrated by the trend to bluer colours at fainter magnitudes.  In
the Burst model of Broadhurst, Ellis \& Shanks (1990) these late-type
systems are associated with star bursting LSB and dwarf galaxies.
Assuming modest clustering growth we derive a correlation length for these
late-type systems of $r_0 \sim 2 h^{-1}$ Mpc.  This is significantly
shorter than the equivalent value for either local bright optically-selected
galaxies, or
IRAS-selected galaxies.  However, it is in reasonable agreement with
the correlation lengths measured for LSB galaxies from the recent SRSS
(Santiago \& da Costa 1990).

We find reasonable support for a simple model of faint galaxy
evolution where luminous galaxies are distributed out to modest
redshifts ($z\sim1$) and the high galaxy surface density is made
up of a dominant co-eval population of bursting LSB galaxies.  The
clustering of this dominant, bursting population can be explained
by plausible theoretical clustering evolution of the locally observed samples.

A simple observational test of our model for the clustering of faint
galaxies would be a correlation analysis of a
deep infra-red ($K$) selected sample, which should show significantly
stronger clustering than we observe.  This results from the dominance,
out to high redshifts, of the same early-type galaxies which define the
local bright optically-selected samples.

\vskip 0.5cm
\noindent{\bf ACKNOWLEDGEMENTS}
\vskip 0.25cm

We acknowledge useful conversations with Gary Bernstein,
Roger Blandford, Warrick Couch, and
Raja Guhathakurta.  We thank Alan Dressler for leading the construction of
the COSMIC instrument. Support under NSF grants AST-89-17765 (TGB) and a
NATO Postdoctoral Fellowship (IRS) is gratefully acknowledged.

\vskip 0.5cm
\noindent{\bf REFERENCES}
\vskip 0.25cm

\noindent Babul, A.\ \& Rees, M.J., 1992, MNRAS, 255, 346.

\noindent Bahcall, J.N.\ \& Soneira, R.M., 1980, ApJS, 47, 357.

\noindent Bardeen, J. M., Bond, J. R., Kaiser, N., \& Szalay, A. S. 1986,
ApJ, 304, 15

\noindent Barrow, J.D., Bhavsar S.P.\ \& Sonoda, D.H., 1984,
MNRAS, 210, 19.

\noindent Bernstein, G.M., Tyson, J.A., Brown, W.R.\ \& Jarvis, J.F.,
1994, ApJ, 426, 516.

\noindent Brainerd, T.G.\ \& Villumsen, J.V., 1994, ApJ, in press.

\noindent Broadhurst, T.J., Ellis, R.S.\ \& Shanks, T., 1988,
MNRAS, 235, 827.

\noindent Broadhurst, T.J., Ellis, R.S.\ \& Glazebrook, K., 1992, Nature,
355, 55.

\noindent Cole, S., Ellis, R., Broadhurst, T.\ \& Colless, M.,
1994, MNRAS, 267, 541.

\noindent Couch, W.J., Jurcevic, J.S.\ \& Boyle, B.J., 1993, MNRAS,
260, 241 (CJB).

\noindent Davis, M., Efstathiou, G., Frenk, C. S., \& White, S. D. M.
1985, ApJ, 292, 371

\noindent Efstathiou, G., Bernstein, G., Katz, N., Tyson, J.A.\ \&
Guhathakurta, P., 1991, ApJ, 380, L47 (EBKTG).

\noindent Fisher, K.B., Davis, M., Strauss, M.A., Yahil, A.\ \& Huchra, J.,
1994, MNRAS, 266, 50.

\noindent Glazebrook, K., Ellis, R.S., Colless, M., Broadhurst, T.J.,
Allington-Smith, J., Tanvir, N.\ \& Taylor, K.\ 1994 MNRAS, in press.

\noindent Groth, E.J.\ \& Peebles, P.J.E., 1977, ApJ, 217, 385.

%
%
%

\noindent Kent, S.M., 1985, PASP, 97, 165.

\noindent King, C.R.\ \& Ellis. R.S., 1985, ApJ, 288, 456.

\noindent Kneib, J.-P., Mathez, G., Fort, B., Mellier, Y., Soucail, G.\ \&
Langaretti, P.-Y., 1994, preprint.

\noindent Koo, D.C.\ \& Szalay, A.S., 1984, ApJ, 282, 390.

\noindent Landy, S.D.\ \& Szalay, A.S., 1993, ApJ, 412, 64.

\noindent Lilly, S.J., Cowie, L.L.\ \& Gardner, J.P., 1991, ApJ, 369, 79.

\noindent Maddox, S.J., Sutherland, W.J., Efstathiou, G., Loveday, J.,
 1990, MNRAS, 234, 692.

\noindent McGaugh, S.S., 1994, Nature, 367, 538.

\noindent Melott, A. 1992, ApJ, 393, L45

\noindent Mould, J.R., Blandford, R.D., Villumsen, J.V., Brainerd,
T.G., Smail, I., Small, T.A.\ \& Kells, W., 1994, MNRAS, submitted.

\noindent Neuschaefer, L.W., Windhorst, R.A.\ \& Dressler, A.,
1991, ApJ, 382, 32 (NWD).

\noindent Peebles, P.J.E., 1980, `The Large-Scale of the Universe',
Princeton Univ.\ Press.

\noindent Guhathakurta, P., Tyson, J.A.\ \& Majewski, S.R., 1990,
ApJ, 357, L9.

\noindent Roche, N., Shanks, T., Metcalfe, N.\ \& Fong, R., 1993,
MNRAS, 263, 360 (RSMF).

\noindent Santiago, B.X.\ \& da Costa, L.N., 1990, 362, 386.

\noindent Saunders, W., Rowan-Robinson, M.\ \& Lawrence, A.,
1992, MNRAS, 258, 134 (SRRL).

\noindent Shanks, T., Fong, R., Ellis, R.S.\ \&
MacGillivray, H.T., 1980, 192, 209.

\noindent Soneira, R.M.\ \& Peebles, P.J.E., 1978, AJ, 83, 845.

\noindent Stevenson, P.R., Shanks, T., Fong, R.\ \& MacGillivray, H.T.,
1985, MNRAS, 213, 953 (SSFM).

\noindent Smail, I., Ellis, R.S., Arag\`on-Salamanca, A., Soucail, G.,
Mellier, Y.\ \& Giraud, E.\ 1993, MNRAS, 263, 628.

\noindent Smail, I., Ellis, R.S.\ \& Fitchett, M.J., 1994, MNRAS, in press.

\noindent Valdes, F.\ 1982, FOCAS Manual, NOAO.

\noindent Yoshii, Y., Peterson, B.A.\ \& Takahara, F., 1993, ApJ, 414, 431.

\vskip 1.5cm
\noindent{\bf FIGURES}
\vskip 0.5cm

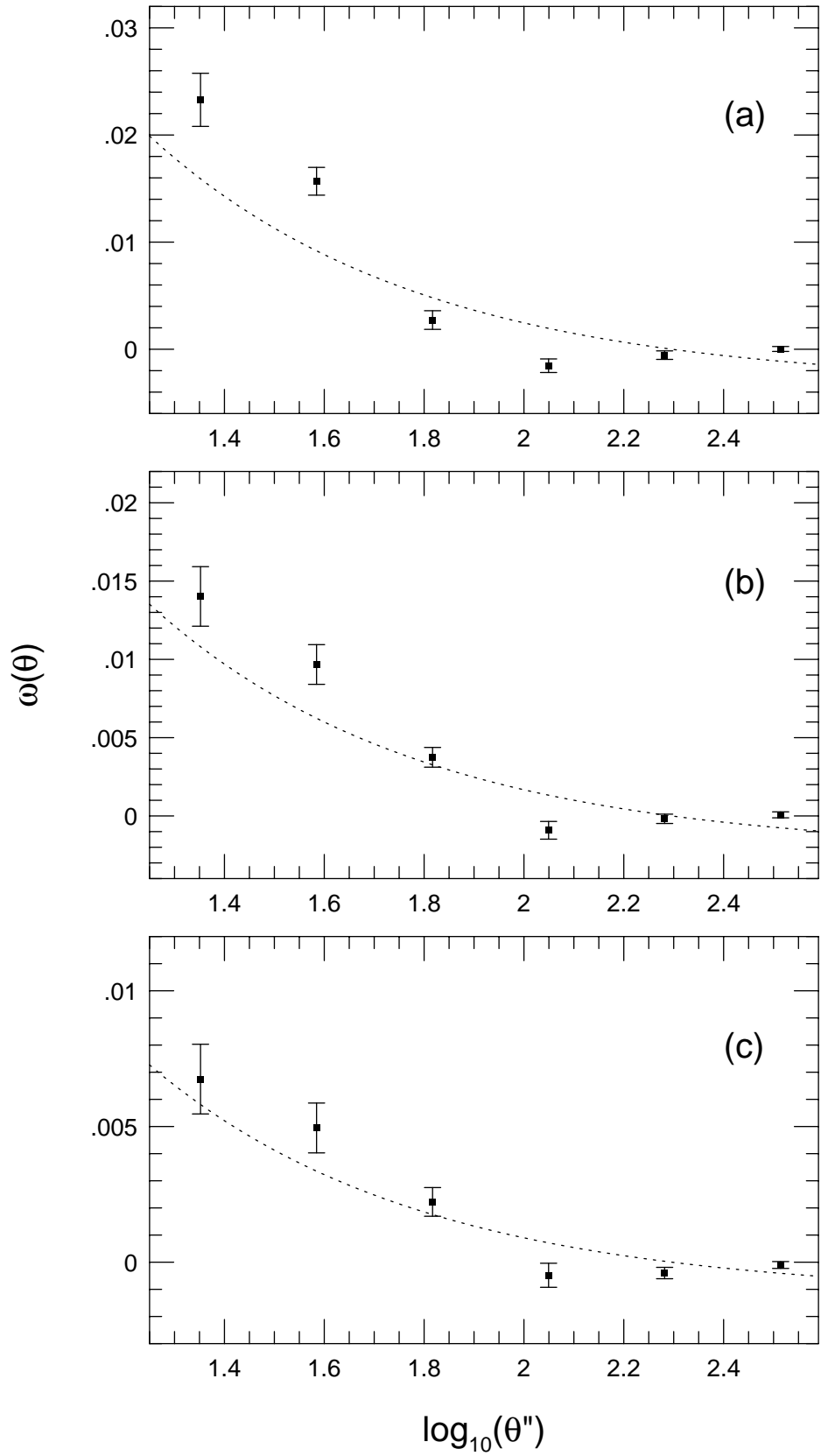
\noindent{\bf Figure~1:}
The two-point angular correlation function of galaxies in 3 magnitude
intervals: (a) $20.0 \le r \le 25.0$, (b) $20.0 \le r \le 25.5$,
(c) $20.0 \le r \le 26.0$).  No correction for the
integral constraint has been made.  Error bars are
estimated from 50 bootstrap resamplings of the data.  The best-fit
power laws of the form $\w = A \theta^{-0.8}$ are indicated by the
dotted lines. Note that different vertical scales are used in each
panel.
\vskip 0.25cm
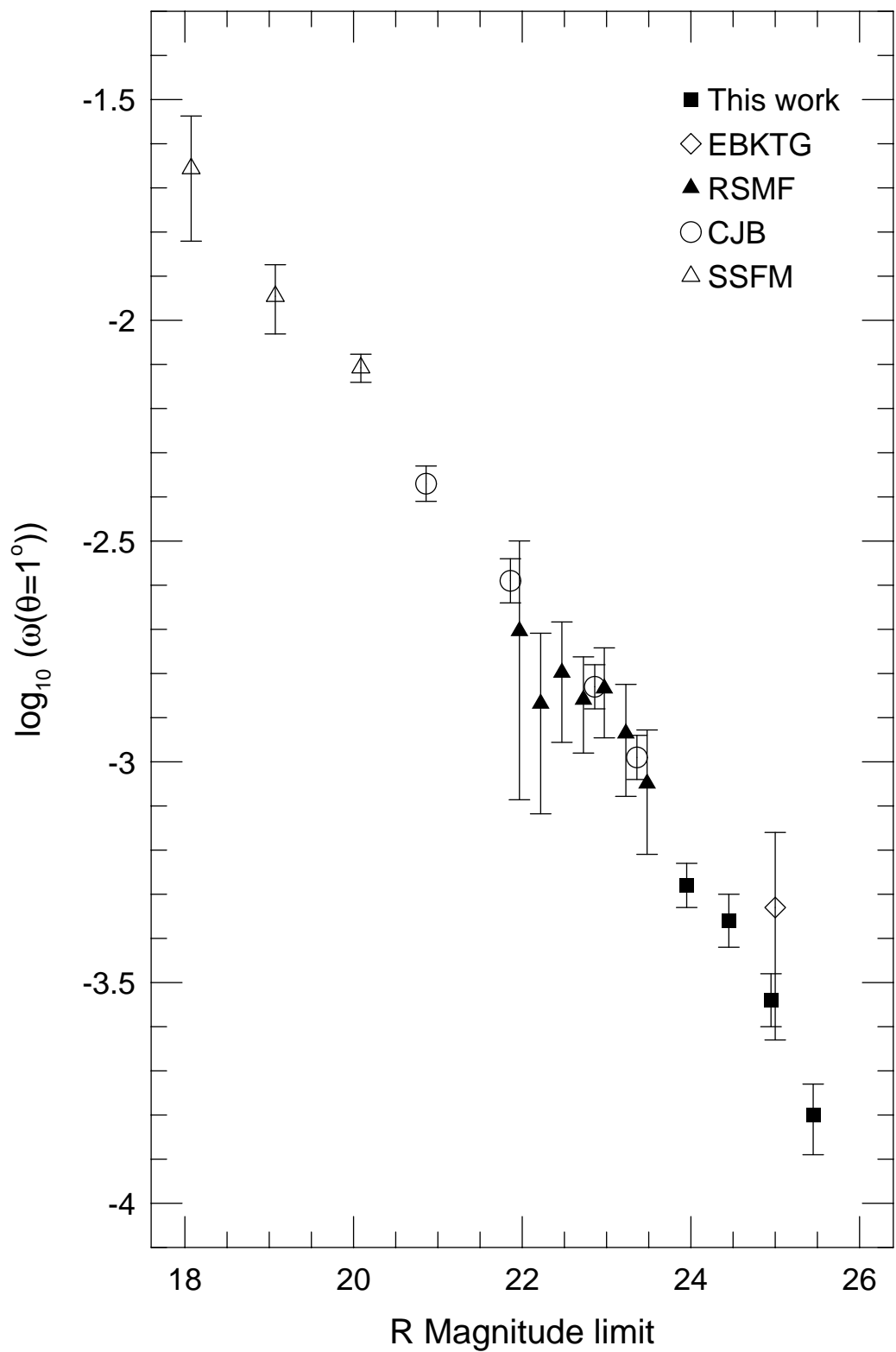
\noindent{\bf Figure 2:}
Corrected $\hat{\omega}(\theta)$ extrapolated 1 degree as function of
limiting $R$ magnitude for a number of red selected samples.
The magnitude limits for the CJB study
have been converted to $R$ using the transformations given by Yoshii,
Peterson \& Takahara (1993) and RSMF.
This figure is adapted from Roche \et (1993).
\vskip 0.25cm
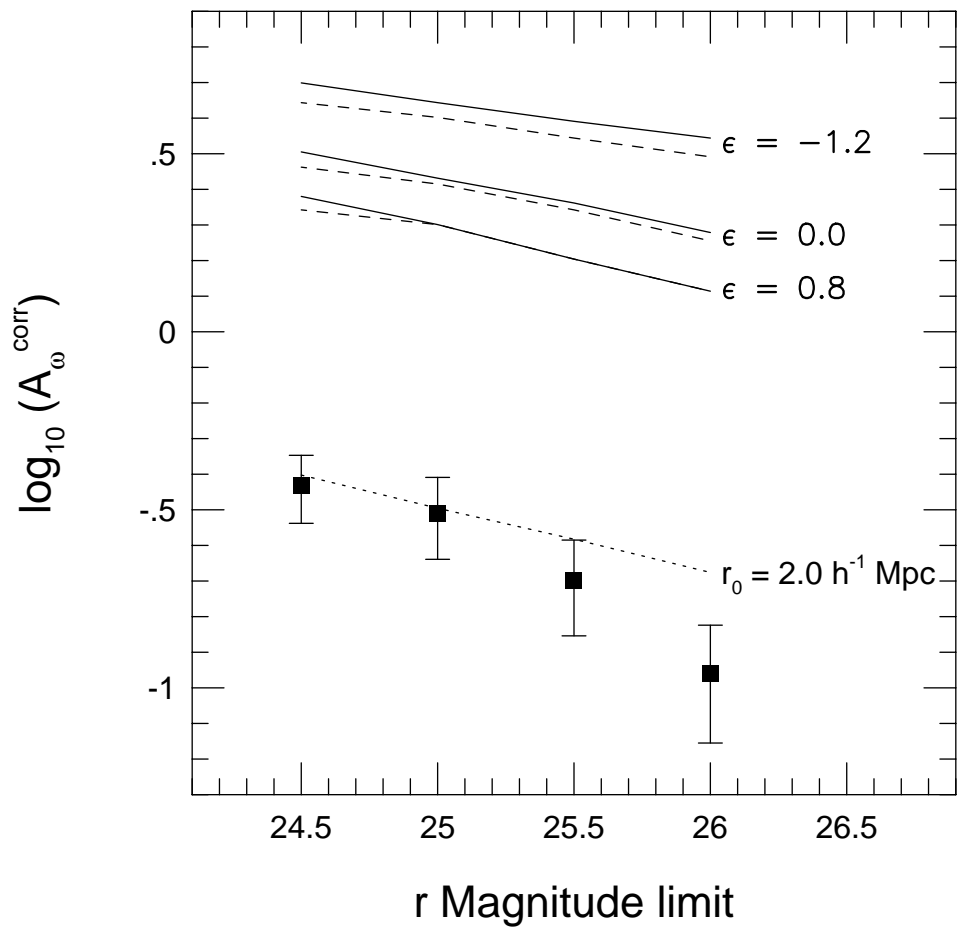
\noindent{\bf Figure 3:}
Observed correlation function amplitude, $A_{\omega}^{\rm corr}$, as a
function of limiting $r$ magnitude (squares).  Error bars indicate
95\% confidence limits.  The predicted amplitude as a function of limiting
magnitude, $\epsilon$, and $\Omega_0$
(assuming $r_0 = 5.5h^{-1}$ Mpc and
$\gamma = 1.8$) is indicated by lines (solid for $\Omega_0 = 1.0$;
dashed for $\Omega_0 = 0.2$).  Also shown (dotted line) is the predicted
$\A^{\rm corr}$ assuming $r_0 = 2.0h^{-1}$ Mpc, $\Omega_0 = 1.0$,
$\gamma = 1.8$, and $\epsilon = 0.8$.

\newpage

\begin{table}
\caption{Observed correlation function parameters}
\begin{tabular}{ccccc}
\hline
\hline
mag. interval & $N_{\rm obs}$ & $\left({N_{\rm obs} \over N_{\rm obs} -
N_{\rm s}} \right)^2$ &
$\A^{\rm corr}$
& $\hat{\omega}(\theta=1^\circ)$ \\
\hline
$20.0 \le r \le 24.5$ & 2170 & 1.439 & 0.37$\pm$0.04 & $(5.3\pm0.6)\times
10^{-4}$ \\
$20.0 \le r \le 25.0$ & 3103 & 1.329 & 0.31$\pm$0.04 & $(4.4\pm0.6)\times
10^{-4}$ \\
$20.0 \le r \le 25.5$ & 4292 & 1.258 & 0.20$\pm$0.03 & $(2.9\pm0.4)\times
10^{-4}$ \\
$20.0 \le r \le 26.0$ & 5730 & 1.206 & 0.11$\pm$0.02 & $(1.6\pm0.3)\times
10^{-4}$ \\ \hline
\end{tabular}
\end{table}

\begin{table}
\caption{Parameters for $N(z)$ assuming a NE model and
corresponding evolution of $\w$}
\begin{tabular}{ccccccccccc}
\hline
\hline
 & & & $\Omega_0 = 1.0$ & & & & & $\Omega_0 = 0.2$ & & \\
\cline{2-5} \cline{7-11}
mag. interval & $z_0$ & $\beta$ & $\epsilon_{\rm stan}$ &
$\epsiras$ & & $z_0$ & $\beta$ &
$\epsilon_{\rm stan}$ & $\epsilon_{\rm IRAS}$ \\ \hline
$20.0 \le r \le 24.5$ & 0.49 & 2.5 & 7 & 3 & & 0.45 & 2.4 & 7 & 3 \\
$20.0 \le r \le 25.0$ & 0.58 & 2.7 & 6 & 3 & & 0.48 & 2.4 & 7 & 3 \\
$20.0 \le r \le 25.5$ & 0.65 & 2.7 & 6 & 3 & & 0.55 & 2.5 & 7 & 3 \\
$20.0 \le r \le 26.0$ & 0.75 & 2.8 & 7 & 4 & & 0.64 & 2.7 & 7 & 4 \\
 \hline
\end{tabular}
\end{table}

\begin{table}
\caption{Shallowest $N(z)$ consistent with the 95\% confidence
level upper limit on $\hat{\omega}(\theta)$}
\begin{tabular}{cccccc}
\hline
\hline
 &  \multispan2{$\Omega_0 = 1.0$}  & &   \multispan2{$\Omega_0 = 0.2$}
  \\ \cline{2-3} \cline{5-6}
mag. interval & $z_0$ & $\beta$ & & $z_0$ & $\beta$  \\ \hline
$20.0 \le r \le 24.5$ & 1.4 & 3  &  & 1.4 & 6   \\
$20.0 \le r \le 25.0$ & 1.8 & 4  &  & 1.6 & 6   \\
$20.0 \le r \le 25.5$ & 2.4 & 6  &  & 1.8 & 6   \\
$20.0 \le r \le 26.0$ & --- & -- &  & 2.5 & 6   \\
 \hline
\end{tabular}
\end{table}

\begin{table}
\caption{Correlation length, $r_0$ ($h^{-1}$ Mpc), required to match
$A_\omega^{\rm corr}$
assuming a NE model $N(z)$ and $\delta = 0.8$ (0.6)}
\begin{tabular}{cccccccc}
\hline
\hline
 & & $\Omega_0 = 1.0$ & & & & $\Omega_0 = 0.2$ & \\ \cline{2-4} \cline{6-8}
mag. interval & $\epsilon=-1.2$ & $\epsilon=0.0$ & $\epsilon=0.8$ &
& $\epsilon=-1.2$ & $\epsilon=0.0$ & $\epsilon=0.8$ \\ \hline
$20.0 \le r \le 24.5$ & 1.4 (1.7) & 1.8 (2.2) & 2.1 (2.6) &
& 1.5 (1.8) & 1.9 (2.3) & 2.2 (2.7)\\
$20.0 \le r \le 25.0$ & 1.4 (1.7) & 1.8 (2.3) & 2.2 (2.8) &
& 1.5 (1.8) & 1.8 (2.3) & 2.2 (2.8) \\
$20.0 \le r \le 25.5$ & 1.2 (1.4) & 1.6 (2.0) & 2.0 (2.4) &
& 1.3 (1.5) & 1.7 (2.0) & 2.0 (2.4) \\
$20.0 \le r \le 26.0$ & 0.9 (1.0) & 1.3 (1.5) & 1.6 (1.9) &
& 1.0 (1.1) & 1.3 (1.5) & 1.6 (1.9) \\
 \hline
\end{tabular}
\end{table}

\end{document}